\documentclass[12pt]{article}
\usepackage{graphicx}
\usepackage{natbib} 
\usepackage{url} 

\usepackage[title]{appendix}
\usepackage{amsmath}
\usepackage{bm}
\usepackage{amsfonts,amssymb} 
\usepackage{algorithm,algorithmic}
\usepackage{lipsum}
\usepackage{cuted}
\usepackage{amssymb}
\usepackage{mathtools}
\usepackage{enumerate}
\usepackage{hhline}
\usepackage{bm}
\usepackage{enumitem}
\usepackage{graphicx}
\usepackage{subfigure}
\usepackage{xcolor}
\usepackage{color, soul}

\usepackage{mathtools}

\DeclarePairedDelimiter\floor{\lfloor}{\rfloor}

\usepackage{lipsum}
\usepackage{cuted}
\usepackage{amssymb}
\usepackage{mathtools}

\newcommand{\blind}{0}

\newtheorem{example}{Example }

\begin{document}

\def\spacingset#1{\renewcommand{\baselinestretch}%
{#1}\small\normalsize} \spacingset{1}


\if0\blind
{
  \title{\bf Sequential Latin Hypercube Design for Two-layer Computer Simulators}
   \author{Yan Wang
   \thanks{Y. Wang  is with the School of Statistics and Data Science, Faculty of Science, Beijing University of Technology, Beijing, China (e-mail: yanwang@bjut.edu.cn)},~
         Dianpeng Wang
        \thanks{\textit{(Corresponding author: Dianpeng Wang)}. D. Wang  is with School of Mathematics and Statistics, Beijing Institute of Technology, Beijing, China (e-mail:wdp@bit.edu.cn)},~
         Xiaowei Yue
 \thanks{X. Yue is with the Grado Department of Industrial and Systems Engineering, Virginia Tech, Blacksburg, VA, 24061 USA (e-mail: xwy@vt.edu)}
 } 
\date{}
  \maketitle
} \fi

\begin{abstract}
The  two-layer computer simulators are commonly used to mimic multi-physics phenomena or systems. 
Usually, the outputs of the first-layer simulator (also called the inner simulator) are partial inputs of the second-layer simulator (also called the outer simulator).
How to design experiments by considering the space-filling properties of inner and outer simulators simultaneously is a significant challenge that has received scant attention in the literature. 
To address this problem, we propose a new sequential optimal Latin hypercube design (LHD) by using the maximin integrating mixed distance criterion. 
A corresponding sequential algorithm for efficiently generating such designs is also developed. 
Numerical simulation results show that the new method can effectively improve the space-filling property of the outer computer inputs. 
The case study about composite structures assembly simulation demonstrates that the proposed method can outperform the benchmark methods.

\end{abstract}

\noindent%
\vfill

\newpage
\spacingset{1.8} 
\section{Introduction}
\label{sec:intro}
{Computer experiments are widely used to emulate physical systems.
 In practice, one engineering system often contains  multi-layer subsystems because of its multi-physics mechanisms or phenomena; the output of one subsystem could be the input for its sequential subsystem. For example, in the earthquake–tsunami model, outputs from the earthquake model are part of the inputs of the tsunami model  \citep{ulrich2019coupled}; the ocean-atmosphere dynamics couples the atmospheric model with the ocean model \citep{nicholls2015impact}; in the  multistage manufacturing processes (MMP), the product quality variations can propagate from one station to its downstream stations  \citep{shi2006stream}. 
 Computer experiments for multi-layer physical systems are challenging due to their high complexity. 
 Two-layer systems exist in large numbers in practice and are also fundamental modules for multi-layer ones.
 Therefore,  this work focuses on two-layer  computer experiments.

 }

\begin{figure}[h!]
\centering
\includegraphics[width=0.85\columnwidth]{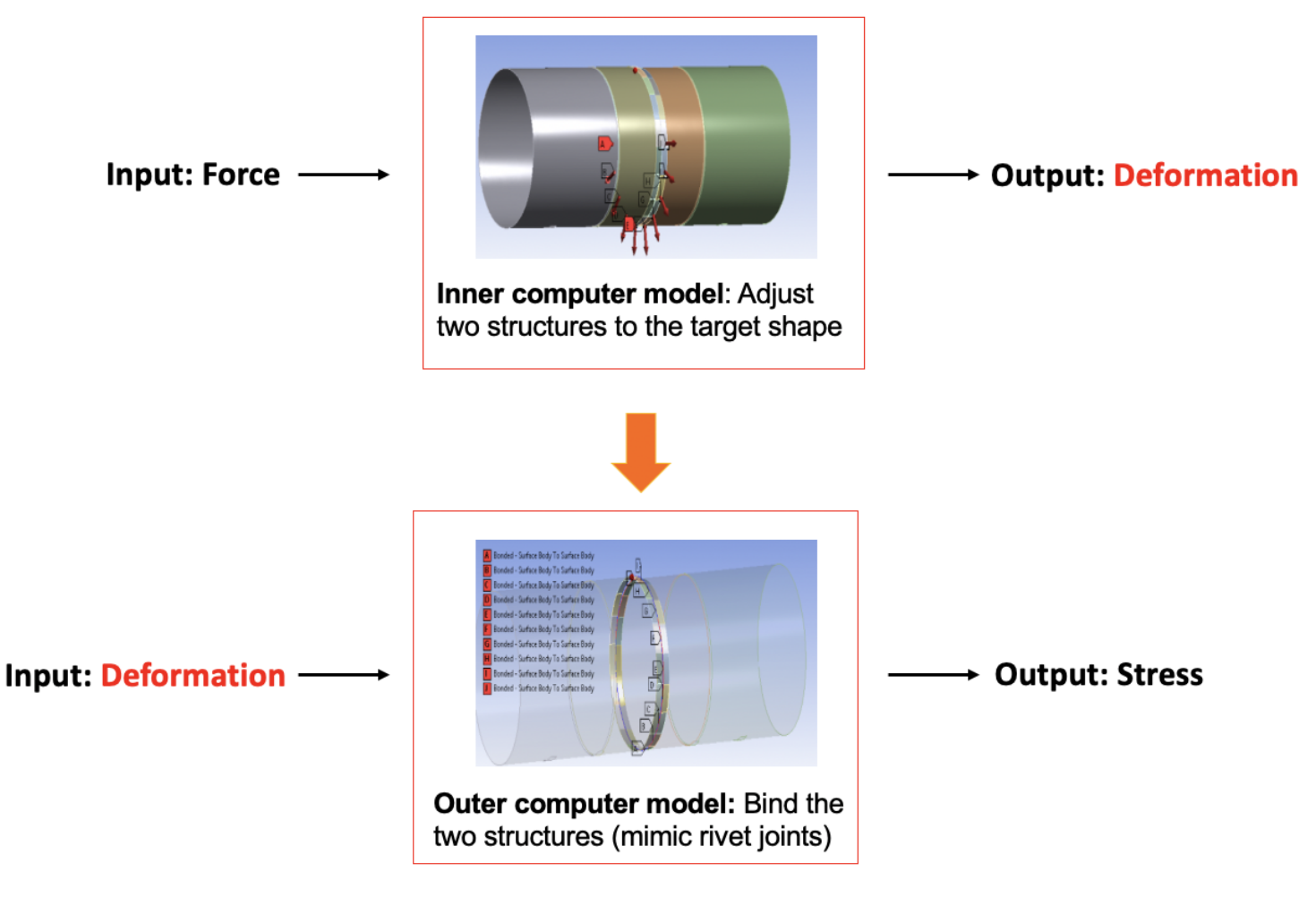}
\caption{The computer experiment mimics the composite structures assembly process.}
\label{Figure flow}
\end{figure}

Composite structure assembly is ubiquitous in many industries, such as aerospace, automotive, energy, construction, etc. To mimic the composite structures assembly process, a computer simulation platform was built based on ANSYS PrepPost Composites workbench \citep{wen2018feasibility,wen2019virtual}, and it has been calibrated via physical experiments \citep{wang2020effective}. 
 This computer simulation can conduct virtual assembly to illustrate detailed composite structure joins, and lay a solid foundation for digital twin simulation of composites assembly. As shown in Figure \ref{Figure flow}, the virtual assembly simulation includes two computer models.
  Specifically, the first layer computer model, named the Automatic Optimal Shape Control (AOSC) system,  simulates the shape control of a single composite structure \citep{yue2018surrogate}. The AOSC can adjust the dimensional deviations of one composite fuselage and make it align well with the other fuselage to be assembled. The second layer computer model simulates the process of composite structures assembly, where the inputs are critical dimensions from two parts, and the outputs are internal stress after assembly \citep{wen2019virtual}. 
 {Obviously, the inputs of the second layer simulator contain the outputs of the first layer simulator.} 
 In practices, the first layer simulator is called the inner computer model, and the second layer one the outer computer model.



\begin{figure}[h]
\begin{center}
\includegraphics[scale=0.6]{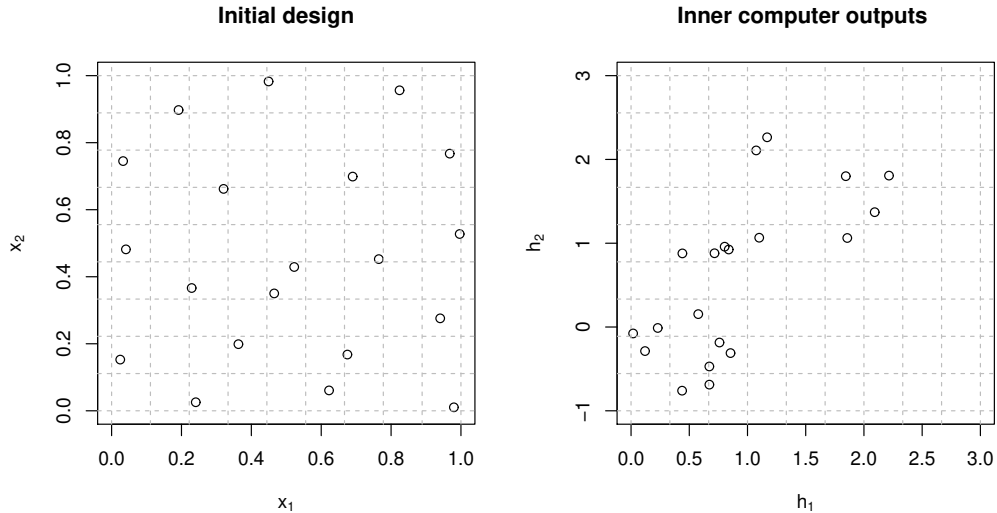}
\end{center}
\caption{Left: Inputs to the inner computer model (MmLHD );  Right: Inner computer outputs. 
 \label{Figure lhd ex}}
\end{figure}

In practice, it is commonly believed that design points should be evenly spread in the experimental space to achieve comprehensive exploration.
Thus, in the literature, the most commonly used experimental design currently for computer experiments is the space-filling design, such as the Latin hypercube design (LHD) \citep{Santner2018}, the maximin Latin hypercube design (MmLHD) \citep{joseph2008orthogonal}, and the rotated sphere packing design \citep{he2019sliced}   etc.
However, for the two-layer computer experiments, the one-shot space-filling design is not appropriate.
Since a one-shot space-filling design means that the space-filling design is applied only to the inputs of the inner computer model. Inputs to the outer computer model are partly determined by the outputs from the inner computer model.
The complexity and non-linearity  of the inner computer model
may yield  uneven inner computer outputs, thus producing a poor design for the outer computer model. 
 As shown in Figure  \ref{Figure lhd ex} {(details about the inner and the outer computer models can be found in Example \ref{ex1}, Section \ref{sec:ex})},  some of the inner computer outputs
 are extremely  close. 
This poor design for the outer computer experiments will result in imprecise uncertainty quantification for the outer computer model. 
Therefore, it is important to develop new designs for two-layer computer experiments.

\cite{kyzyurova2018coupling} adopted the space-filling designs for the inner and outer computer experiments separately. This method treats the outputs of the inner computer trial as control variables for the outer trial design, which has three problems: (1) the design region for these control variables is difficult to determine; (2) the outer computer trial ignores the information provided by the inner computer trial; and (3) the method is only applicable to the case where the inner and outer computer trials can be performed independently. For the two-layer computer trials that cannot be performed independently, the scheme cannot be implemented. 
\cite{marque2019efficient} and
\cite{ming2021linked} adaptively generated a sequential optimal design for a  two-layer computer model based on a nested Gaussian process model, which assumes that both the inner and outer computer experiments are implementations of the Gaussian process. This kind of approaches are highly dependent on the Gaussian assumptions made about the inner and the outer computer model.

In this work, in order to deal with the challenge of experiments for the  two-layer computer experiments, we derive a new distance metric for the inputs of the outer simulator and propose an efficient sequential design scheme by maximizing the minimum new distance between the inputs of the outer simulator.
The proposed design can consider the space-filling property of both the inputs and the outputs for the inner computer experiments. Our contributions can be summarized as follows: 
 \begin{itemize}
  \item   A new distance named the \emph{Mixed-distance} is introduced to measure the inter-site distance between the inputs of the outer computer simulator. This distance is a weighted average of the inter-site distance between the inner computer inputs and the inter-site distance between the inner computer outputs.
   \item  For the multi-response inner computer model,  principal component analysis is adopted to reduce the dimensionality of the inner computer outputs.
   Independent Gaussian Process models \citep{Santner2018} are built to mimic the principal component scores. The distance between the principal component scores is used to measure the distance between the inner computer outputs. The closed form for this distance can be derived. 
   \item {  
 Conventional space-filling design algorithms with deterministic distances are not sufficient when the inputs of the outer model are unknown. To overcome this challenge, the proposed distance between the inner computer outputs considers the uncertainty of the surrogate models. The proposed design can dive into the internal nested structure and achieve the optimal space-filling properties for both the inner computer model and the outer computer model. }
    \item The Maximin design criterion and a fast algorithm are proposed to generate sequential designs for the  two-layer computer experiments. 
    The performance of the proposed method is shown in numerical studies. The proposed method is applied to generate designs for the composite structures assembly experiments.

\end{itemize}

The outline of this paper is as follows. Section \ref{sec:nce} introduces two-layer computer experiments.  Section \ref{sec:cri} presents the proposed sequential design and gives the detailed algorithm. Section \ref{sec:ex} examines the performance of the proposed method through two numerical studies. 
In Section \ref{sec:case}, the proposed method is applied to generate the design for  the composite structures assembly simulation.  Concluding remarks and the discussions are given in Section \ref{sec.dis}.

\section{ Two-layer Computer Experiments}
\label{sec:nce}
Denote $f:[0,1]^K\rightarrow\mathbb{R}^M$ to be a deterministic  two-layer computer model, which is defined as 
\begin{equation}
\label{ncm-1}
f(\bm x)=\bm g\left(
\begin{bmatrix} \bm x  \\ \bm h(\bm x) \end{bmatrix}  
\right),\bm x \in [0,1]^K,
\end{equation}
where $\bm h(\bm x)=[h_1(\bm x),\ldots,h_L(\bm x)]^T$ is the $L\times 1$ vector of the inner computer outputs, 
and $\bm g\left(\bm x^{outer}  
\right)=\left[g_1\left(\bm x^{outer}  
\right),\ldots,g_M\left(\bm x^{outer}  
\right)\right]^T$ is the $M\times 1$ vector of the  outer computer outputs,  with $\bm x^{outer}=[\bm x^T, \bm h^T(\bm x) ]^T\in [0,1]^K\times \mathbb{R}^{L}$  an input of the outer computer  model.
Here, we assume the outputs for both the inner and outer computer models are scalar. In fact, in many cases, the functional outputs can be discretized to scalar outputs \citep{chen2021function}. It is worth noticing that the design variables in the model (\ref{ncm-1}) are the combinations of all control variables. The  two-layer computer model (\ref{ncm-1}) contains the following cases: (1) $f(\bm x)=\bm g(\bm h(\bm x))$, in which $\bm g$ depends on $\bm x$ only through $\bm h$; (2) $\bm x$ can be divided into two parts, i.e., $\bm x=(\bm x_1,\bm x_2)$ and $f(\bm x)=\bm g\left([\bm x^T_2, \bm h^T(\bm x_1)]^T \right)$ (or $f(\bm x)=\bm g\left([\bm x^T, \bm h^T(\bm x_1) ]^T\right)$) as shown in \cite{marque2019efficient} and \cite{wang2021nested}. In these cases, the inner computer model $\bm h(\cdot)$ is  insensitive to part of the input variables.

\begin{figure}[h]
\begin{center}
\includegraphics[scale=0.3]{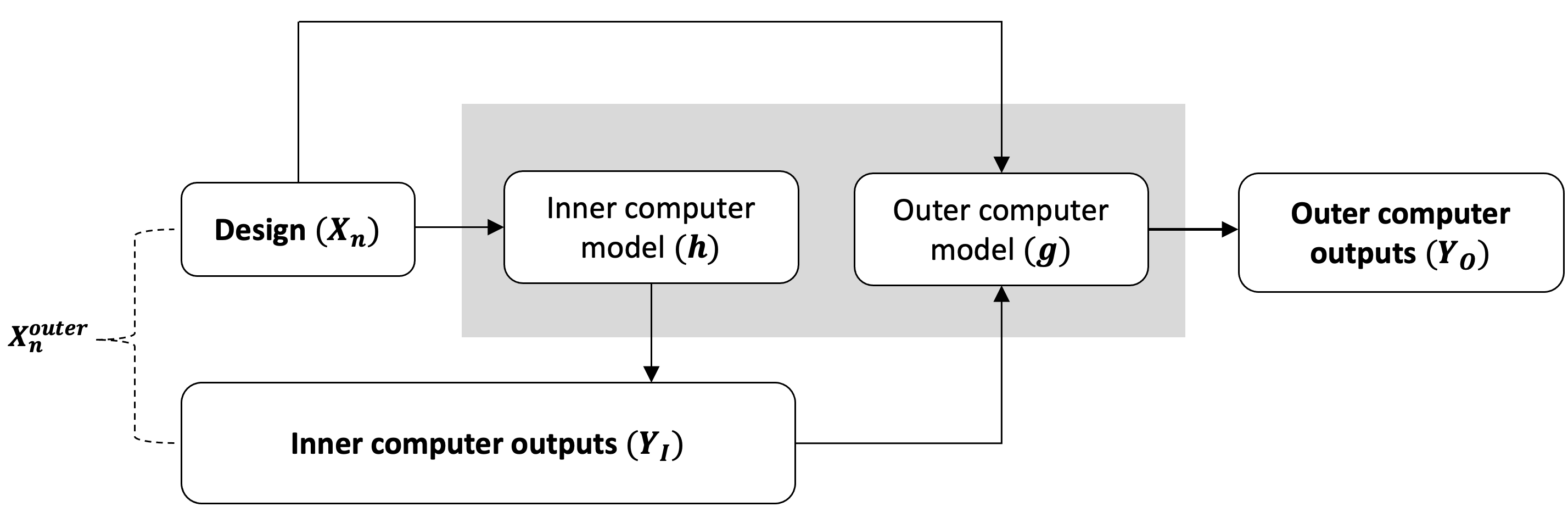}
\end{center}
\caption{Two-layer computer experiments.  
 \label{Figure nce}}
\end{figure}

Suppose $\bm X_n=\{\bm x_1,\ldots,\bm x_n\}$ is a $n$-point experiment design, with $\bm x_i\in [0,1]^K$. Let $\bm H_l=[h_l(\bm x_1),\ldots,h_l(\bm x_n)]^T$, $l=1,\ldots,L$ denote the $n\times 1$ vector of the corresponding responses of $h_l(\cdot)$, and $\bm Y_I=[\bm H_1,\ldots, \bm H_L]$ denote  the $n\times L$ matrix of the inner computer outputs.  Let $\bm X_n^{outer}=\{\bm x^{outer}_i\}_{i=1}^n$ denote the set of inputs of the outer computer model and $\bm G_{m}=\left[g_m\left(\bm x_1^{outer}  
\right) ,\ldots, g_m\left(
\bm x_n^{outer} 
\right)\right]^T$, $m=1,\ldots,M$  the $n\times 1$ vector of the  corresponding $m$-th outer computer outputs and $\bm Y_O=[\bm G_1,\ldots, \bm G_M]$ denote  the $n\times M$ matrix of the outer computer outputs.
Figure  \ref{Figure nce} shows the framework of  two-layer computer experiments.

 Assume that { both the inner and the outer computer models are deterministic but expensive to evaluate. The `deterministic' here refers to that the same input will deduce the same computer output.} As discussed above, a one-shot space-filling design $\bm X_ n$ may yield uneven inputs to the outer computer model. To fully explore, optimize or gain insight into the system, 
 the 
space-filling properties of both the inner computer experimental design and the outer computer inputs need to be considered. 
That is, 
an essential consideration in the design for the  two-layer computer experiments is the trade-off between \emph{exploration} ( $\bm X_ n$ must be space-filling to a certain degree) and \emph{exploitation} (  $\bm X_n^{outer}$  should fill up the interesting domain of the outer computer model).
Sequential design methods can solve this problem by iteratively selecting design points. 
Because the space-filling property of  $\bm X_n^{outer}$ depends on the complexity of the inner computer model,  the proposed sequential strategy updates the design by ``learning''   the inner computer model and assessing the quality of  $\bm X_n^{outer}$ adaptively.


\section{Sequential Mixed-distance Design }
\label{sec:cri}

To assess the quality of $\bm X_n^{outer}$,  {a new mixed} inter-site distance is introduced in Subsection \ref{subsec:ocm}. This distance takes into account both the inter-site distance between the inner computer outputs and the inter-site distance between the design variables.
In Subsection \ref{subsec:cri}, we introduce the maximum criterion and apply this criterion to query the points sequentially. 
\subsection{{The Mixed-distance } }
\label{subsec:ocm}

 Recall that $\bm x^{outer}=\left(\bm x,\bm h(\bm x)\right)\in [0,1]^K\times \mathbb{R}^{L}$ is an input of the outer computer  model. 
  Define $d^{outer}_{i,j}$ to be the  inter-site distance  between $\bm x_i^{outer}$ and $\bm x_j^{outer}$ :
 \begin{equation}
 \label{dis}
d^{outer}_{ij}=d(\bm x_i^{outer},\bm x_j^{outer})=wd(\bm h_i,\bm h_j)+(1-w)d(\bm x_i,\bm x_j),
  \end{equation}
  where $w\in [0,1]$ is a   pre-specified  weight;  $d(\bm h_i,\bm h_j)$ measures the distance between $\bm h(\bm x_i)$ and $\bm h(\bm x_j)$, which is defined later; 
{$d(\bm x_i,\bm x_j)=\left[\sum_{k=1}^K|x_{i,k}-x_{j,k}|^2\right]^{1/2}$ is the  Euclidean  distance between $\bm x_i$ and $\bm x_j$.} {Obviously, this distance is a weighted mixture of the inter-site distance of the inner computer inputs and outputs. We call this distance the \emph{Mixed-distance}.}


 Next,  we  elaborate on the definition of $d(\bm h_i,\bm h_j)$. { Since the inner computer model is expensive,  only a limited number of inner computer experiments can be run. Hence, inputs $\bm h(\bm x)$ to the outer model are unknown if it is not evaluated at $\bm x$. To estimate $\bm h(\bm x)$, a surrogate model can be built to mimic the  inner computer model. 
 In practice, the GP model is always used to build the surrogate model, which will lead to uncertainty about the unobserved $\bm h(\bm x)$.
 Thus, the conventional (deterministic) distances are not suitable here because they ignore the uncertainty of the surrogate model.}
 From  \cite{ortega2021stochastic}, the  {$L_2$ metric  
$\left\{{\rm E}
\left[\sum_{l=1}^L|h_l(\bm x_i)-  h_l(\bm x_j)|^2\right]\right\}^{1/2}$} is a common choice to measure the distance between two random vectors. 
 Since $h_{l}$ and $h_{l'}$, $l\neq l'\in {1,\ldots, L}$ may be dependent,  
 it requires the marginal distribution functions of $h_l, l=1,\ldots,L$ to evaluate this distance. 
 Deriving these marginal distribution functions is a  difficult task.  To simplify the calculation of the distance, 
we perform a principal component analysis (PCA) by utilizing the singular value decomposition (SVD) of the scaled $\bm Y^T_I$.  
Then we replace $d(\bm h_i,\bm h_j)$ by the distance between the first $L^{pc}$ independent principal component scores (PCs). 
Such measures have been used in several published studies, for example, \cite{ding2002adaptive}.  
In this work, we assume that $L$ is not too large. Because in general, when $L$ is relatively large, the finite inner computer outputs will not be overlapped or be very close to each other.
Even though, another advantage of PCA is  reducing the dimension of $\bm h(\cdot)$, thus further simplifying the calculation.

Let $\bm Y^{pc}_I=[\bm H^{pc}_1,\ldots,\bm H^{pc}_{L^{pc}}]$  with  $\bm H^{pc}_l=[h^{pc}_l(\bm x_1),\ldots,h^{pc}_l(\bm x_n) ]^T$, $l=1,\ldots, L^{pc}$ denote the $L^{pc}$ uncorrelated PCs,  and $ L^{pc}<\min (L,n)$. Suppose $h_l^{pc}$ is one realization of a GP.
Measure the distance between $\bm h(\bm x_i)$ and $\bm h(\bm x_j)$ by using the following {$L_2 $ metric:
  \begin{equation}
 \label{dist-h}
 d(\bm h_i,\bm h_j)=\left\{{\rm E}\left[\sum_{l=1}^{L^{pc}} |  h^{pc}_l(\bm x_i)-  h^{pc}_l(\bm x_j)|^2\right]\right\}^{1/2}.
   \end{equation}}
Since  $d(\bm x_i,\bm x_j)$ and $d(\bm h_i,\bm h_j)$ are both valid $L_2$ metrics, the Mixed-distance  (\ref{dis}) is a valid metric.
In addition,  for the PCs from the PCA procedures, the following conclusions hold \citep{jolliffe2002principal}:
(I) the Euclidean distance calculated using all $L$ PCs is identical to that calculated from the scaled $\bm Y^T_I$; (II) suppose that $L^{pc}$($<L$) PCs account for most of the variation in the scaled $\bm Y^T_I$. Using $L^{pc}$
PCs provide an approximation to the original Euclidean
distance. As the original distance is calculated from the scaled $\bm Y^T_I$, the distance  (\ref{dist-h}) used in this work  set equal weights on each $h_l^{pc}$. That is, we don't scale the $h_l^{pc}$ to $[0,1]$ as \cite{joseph2008orthogonal}.

Next, we focus on the calculation of $d(\bm h_i,\bm h_j)$. By the assumptions on $h_l^{pc},l=1,\ldots, L^{pc}$, independent GP models can be adapted to mimic the PCs. Details about the GP modeling can be found in  Appendix \ref{emu-gp}. Denote the posterior mean and variance functions of $h^{pc}_l (\bm x_i)-h^{pc}_l(\bm x_j)$, i.e., mean and variance
functions of $ h^{pc}_l (\bm x_i)-h^{pc}_l(\bm x_j)$ conditioned on $\bm X_n, \bm Y_I$, by $\mu_l(\bm x_i,\bm x_j)$ and $\tau^2_{l}(\bm x_i,\bm x_j)$, respectively.  
{Based on the moments of normal distribution \citep{winkelbauer2012moments}, it can be easily verified that 
the closed form solution of $d(\bm h_i,\bm h_j)$ is
  \begin{equation}
  \label{cf2}
 \begin{aligned}
d^2(\bm h_i,\bm h_j)
=&\sum_{l=1}^{L^{pc}} (\mu_l^2+\tau_l^2).
\end{aligned}
  \end{equation}}
Specifically, for any $\bm x\in [0,1]^K$, 
there is $|\mu_l(\bm x,\bm x_i)|=|\hat {h}^{pc}_{l}( \bm x)-{ h}^{pc}_{l}( \bm x_i)|$ and $\tau_l=s_l(\bm x)$. Here, $\hat {h}^{pc}_{l}( \bm x)$  is the GP predictor of ${ h}^{pc}_{l}$ at the point $\bm x$ and $s_l(\bm x)$  the standard deviation. The specific forms of $\hat {h}^{pc}_{l}( \bm x)$ and $s_l(\bm x)$  can be found in (\ref{blue}) and (\ref{var}), respectively.  
Suppose  $\hat {  h}^{pc}_{l}( \bm x), l=1,\ldots, L^{pc}$ are a continuous function and differentiable on $[0,1]^K$. { This assumption can be guaranteed by chosen Gaussian correlation functions (\ref{exp}) and Mat\'ern correlation functions (\ref{matern}) as  the correlation functions.} From one-order Taylor expansion of $h^{pc}_l(\bm x)$ at $\bm x_i, i\in \{1,\ldots,n\}$, we have  that, 
$$|\hat {h}^{pc}_{l}( \bm x)-{ h}^{pc}_{l}( \bm x_i)|= \left|\frac{\partial \hat {  h}^{pc}_{l}( \bm x)}{\partial \bm x}\right|| \bm x- \bm x_i|+O(| \bm x- \bm x_i|^2).$$

{ Combining with the definition of the Mixed-distance  (\ref{dis}), we have that} larger distance between $\bm x$ and $\bm x_i$, $i=1,\ldots,n$, larger absolute value of $\frac{\partial \hat {  h}^{pc}_{l}( \bm x)}{\partial \bm x}$ and  larger $s_l(\bm x)$ will deduce to larger Mixed-distance. A large absolute value of   $\frac{\partial \hat {  h}^{pc}_{l}( \bm x)}{\partial \bm x}$ implies that the  inner GP predictor varies drastically; a large $s_l(\bm x)$ implies a large uncertainty of the GP predictor for the $l$-th PCs. 

 { It is worth noticing that, a simpler alternative of the definition of $d(\bm h_i,\bm h_j)$ is a standard deterministic distance between just the posterior mean of  the $L^{pc}$ PCs. That is,  for any unobserved $\bm x\in [0,1]^K$, there is  $d^2\left(\bm h_i,\bm h(\bm x)\right)
=\sum_{l=1}^{L^{pc}} \left[\hat {h}^{pc}_{l}( \bm x)-{ h}^{pc}_{l}( \bm x_i)\right]^2.$  Relative to  this deterministic distance,  the proposed one (\ref{cf2}) considers the variance/uncertainty in the GP predictors for the PCs. } 

\subsection{Maximin-distance design criterion}
\label{subsec:cri}
   \cite{johnson1990minimax} showed that the maximin  distance design is asymptotically optimum under a Bayesian setting.  Moreover, this criterion optimizes the worst case, thus generating robust space-filling designs.  We adopt the maximin distance criterion to
 query a sequential point, that is,
  \begin{equation}
  \label{Mm}
  \begin{aligned}
\bm x_{n+1}=\max_{\bm x\in [0,1]^K}\min _{1\leq i\leq n}wd(\bm h_i,\bm h(\bm x))+(1-w)d(\bm x_i,\bm x).
\end{aligned}
  \end{equation} 
 { The obtained sequential design is called \emph{sequential mixed-distance design}, abbreviated as SMDD.}
  The weight  $w$ balances  the space-filling property  of the two parts of outer computer inputs. If $w=0$, $\bm x_{n+1}$ is a sequential point that maximizes the minimal distance between $\bm x$ and $\bm x_i$. If $w=1$,  the maximin distance property of $\bm h_{n+1}$ is guaranteed, whereas the maximin distance property of $\bm x_{n+1}$ is lost. The choice of $w$ is a hard task because $d(\bm h_i,\bm h(\bm x))$ and $d(\bm x_i,\bm x)$ have different scales. 

   A natural idea to avoid determining the weights  is that we first pre-specify a set of candidate points, whose space-filling properties are guaranteed, and then obtain the sequential points by maximizing the minimum of $d(\bm h_i,\bm h(\bm x))$, that is,
    \begin{equation}
  \label{Mm2}
  \begin{aligned}
\bm x_{n+1}=\max_{\bm x\in X_{cand}}\min _{1\leq i\leq n}d(\bm h_i,\bm h(\bm x)),
\end{aligned}
  \end{equation} 
where $X_{cand}$ is the set of candidate points. 
The choice of candidate points $X_{cand}$ affects the accuracy and the efficiency of the optimization (\ref{Mm2}).  {Choosing points sequentially from a larger candidate set based on the maximin-design criterion  (\ref{Mm2}) is commonly referred to as the computer-aided design of experiment (CADEX) algorithm \citep{kennard1969computer}.}
 A very straightforward choice for $X_{cand}$ is an LHD. To avoid sequentially adding points too close to existing points,  we suggest using the sliced LHD (SLHD), such as sliced MmLHD  \citep{ba2015}, balanced sliced orthogonal arrays \citep{ai2014construction}, the flexible sliced designs \citep{kong2018flexible} and the sliced rotated sphere packing designs \citep{he2019sliced} etc. to generate the initial and the candidate points. 

Because many different designs may have the same minimum inter-site distance, an extension of the maximin criterion is the $\phi_q$ criterion. This $\phi_q$ criterion minimizes the average reciprocal  interpoint distance between the input points \citep{jin2005efficient}. Then   (\ref{Mm2}) becomes
 \begin{equation} 
 \label{Mm3}
  \begin{aligned}
\bm x_{n+1}=\min_{\bm x\in X_{cand}} \phi_q,
\end{aligned}
  \end{equation} 
  where 
 \begin{equation}
\label{phiq}
\phi_q(\bm x)=\left\{\sum_{1\leq i \leq n} d(\bm h_i,\bm h(\bm x))^{-q}\right\}^{1/q}, q>0.
  \end{equation}
 For large enough $q$, minimizing $\phi_q$ is equivalent to maximizing the minimum distance
among the design points.  



 Let $n_0$ be the sample size of the initial design and $N$ be the sample size of the final sequential design.
 An usually used stopping rule in the literature is the number of runs.
 Assume the sample size of the candidate points, which are used to choose the sequential design point is $N_c$.
 When the dimension of $\bm x$ is large, it is reasonable to use more runs.
 This requires a large number of candidate points.
 Thus we suggest setting $N_c=a(N-n_0)K$, where $a$ is a pre-specified constant.
 The corresponding procedure for generating the desired design can be described in Algorithm \ref{alg:1}. 
\begin{algorithm}
	\caption{Sequential LHD}\label{alg:1}
  \begin{algorithmic}[0]
		\STATE  Obtain an initial design $\bm X_{n_0}$ with $n_0$ points, and run the inner computer simulator  at these points,
yielding the corresponding simulator outputs $\bm Y_I$;
   \STATE Generate $N_c$ candidate samples, denoted as  $\mathcal{D}_c$;
         \FOR{ iteration $n=n_0,\cdots,N-1$}
         
         \STATE Centre the matrix $\bm Y^T_I$ and denote as $\bm Y^{*T}_I$, so that the mean of
each column of $\bm Y^{*T}_I$ is zero, and the variance of each column is one;
        \STATE Calculate the singular value decomposition of $\bm Y^{*T}_I$, i.e., $\bm Y^{*T}_I = S V\Gamma^T$, where $\Gamma$ is an $L\times L$  unitary matrix containing the principal components (the eigenvectors), $V$ is an $n\times L$ diagonal matrix containing the ordered principal values (the eigenvalues), $S$ is an $n\times n$ matrix containing the left singular values;
        
        \STATE Choose $L^{pc}$ such that the cumulative variation of the first $L^{pc}$ PCs is more than $90\%$ of total variation;
        
        \STATE Project $\bm Y^{*T}_I$ onto the principal subspace which is denoted as $[\bm H^{pc}_1,\ldots,\bm H^{pc}_{L^{pc}}]$, where  $\bm H^{pc}_l=[h^{pc}_l(\bm x_1),\ldots,h^{pc}_l(\bm x_n) ]^T$, $l=1,\ldots, L^{pc}$

  \STATE Build independent GP models to mimic $[\bm H^{pc}_1,\ldots,\bm H^{pc}_{L^{pc}}]$;
        \STATE
    Find $\bm x_{n+1}=\min_{\bm x\in \mathcal{D}_c}\phi_q(\bm x) $;
  
          \STATE     
         Run the inner computer code at  $\bm x_{n+1}$ , augment $\bm X_N$ and $\bm Y_I$ with  $\bm x_{n+1}$ and $\bm h(\bm x_{n+1})$;   
         \ENDFOR

         \RETURN $\bm X_N$ and $\bm Y_I$;  \end{algorithmic}
\end{algorithm}
  


Note that, the PCs are obtained from the sample  matrix $\bm Y_I$. As the iteration proceeds,   the sample size of $\bm Y_I$ increases, and the reasonable value for $L^{pc}$ could generally become larger  \citep{jolliffe2002principal}. 
To ensure the stability of outcomes of principal components analysis,  the ratio $n_0/L$ should be larger than $3$ \citep{grossman1991principal}. 
Considering the accuracy of GP models, we suggest the initial sample size $n_0=\max(n_0^*,3L)$, where  $n_0^*$ is the sample size that  can guarantee the accuracy of the GP models. 
Besides, if the inner computer outputs  can be mimicked by independent GP models, then the PCA can be omitted. $n_0$ is determined by $n_0^*$.
The common choice of $n_0^*$ is $10K$, as recommended in \cite{Loeppky2009Special}.




\section{Numerical Simulations}
\label{sec:ex}
{ In this section, we compare the performance of three methods: (1) SMDD: the proposed sequential mixed-distance design;  (2) SMDD-Det: sequential mixed-distance design when $d(\bm h_i,\bm h_j)$ is the standard deterministic distance between the $\hat {h}^{pc}_{l}( \bm x)$ value;  (3) MmLHD: maximin Latin hypercube design with the sample size equals to $N$. 
MmLHD does not depend on the initial design. 
The following two initial designs are considered for the SMDD and SMDD-Det.
\begin{itemize}
    \item ${\rm{ID}}_1$: The initial design $\bm X_{n_0}$ is an MmLHD;
    \item ${\rm{ID}}_2$: Poorly space-filled initial design, e.g., some parts of the design region are not explored.
\end{itemize}
Here, by comparing the performance of the first and the second method, we can explore the advantage of introducing the distance (\ref{dist-h}).  Involving two  initial designs can help us
further understand the impact of the initial design and the accuracy of the inner GP models on the sequential designs.

The works of \cite{marque2019efficient} and  \cite{ming2021linked}  assumed that both the inner and outer computer experiments are realizations of stationary GPs. 
{A stationary GP assumes that the function of interest has the same degree of smoothness in the whole covariate space. This assumption is too strong in many cases \citep{wang2021nested}. 
Taking composite structure (aircraft) assembly as an example, the outer computer simulator describes the riveting process, which could be influenced by dynamic forces, varying input shape deviations, and generated residual stresses. For more details refer to the case study.}

{To obtain a design robust to the model assumption for the outer computer simulators,  different from the optimal design \citep{marque2019efficient,ming2021linked}, the proposed method focuses on the space-filling property of the design for the outer computer experiments.} 
The average inter-site distances (abbreviated as \emph{AID}) of $\bm X_N$ and $\bm Y_I$ are used to compare the performance of different designs, which are defined as 
  \begin {equation*}
    \begin{aligned}
{\rm{AID}}_x&=\frac{2}{N(N-1)}\sum_{i=1,i\leq j}^N\sum_{j=1}^N d(\bm x_i,\bm x_j), \\
{\rm{AID}}_h &=\frac{2}{N(N-1)}\sum_{i=1,i\leq j}^N\sum_{j=1}^N d( \bm h_i, \bm h_j),
  \end{aligned}
  \end {equation*}
It is obvious that the design with the larger AID is better for the same sample size.

For a fair comparison,  we started the iteration with the initial sample size $n_0=\max(10K,3L)$. The stopping rule is set on the sample size of the final sequential design $N$. Set  $N_c=5(N-n_0)K$, the $N_c$ candidate points are generated from the sliced MmLHD  \citep{ba2015}.  The sequentially added point is then queried by minimizing $\phi_q(\bm x)$ over the candidate points. Here $q$ is set on $15$ \citep{jin2005efficient}. } 

\begin{example}
\label{ex1}
Suppose the inner computer simulator includes the three-hump Camel function and six-hump Camel function:
  \begin {equation*}
  \begin{aligned}
h_1(\bm x)&=2\bar x_1^2-1.05\bar x_1^4+\frac{\bar x_1^6}{6}+\bar x_1\bar x_2+\bar x_2^2,\\
h_2(\bm x)&=(4-2.1\bar x_1^2+\frac{\bar x_1^4}{3})\bar x_1^2+\bar x_1\bar x_2+(-4+4\bar x_2^2)\bar x_2^2,
\end{aligned}
  \end{equation*}
  where $\bar x_1=2x_1-1$, $\bar x_2=2x_2-1$, $\bm x=(x_1,x_2)\in [0,1]^2$.
  \end{example}      
    \begin{figure}[htbp]
\begin{center}
\includegraphics[scale=0.6]{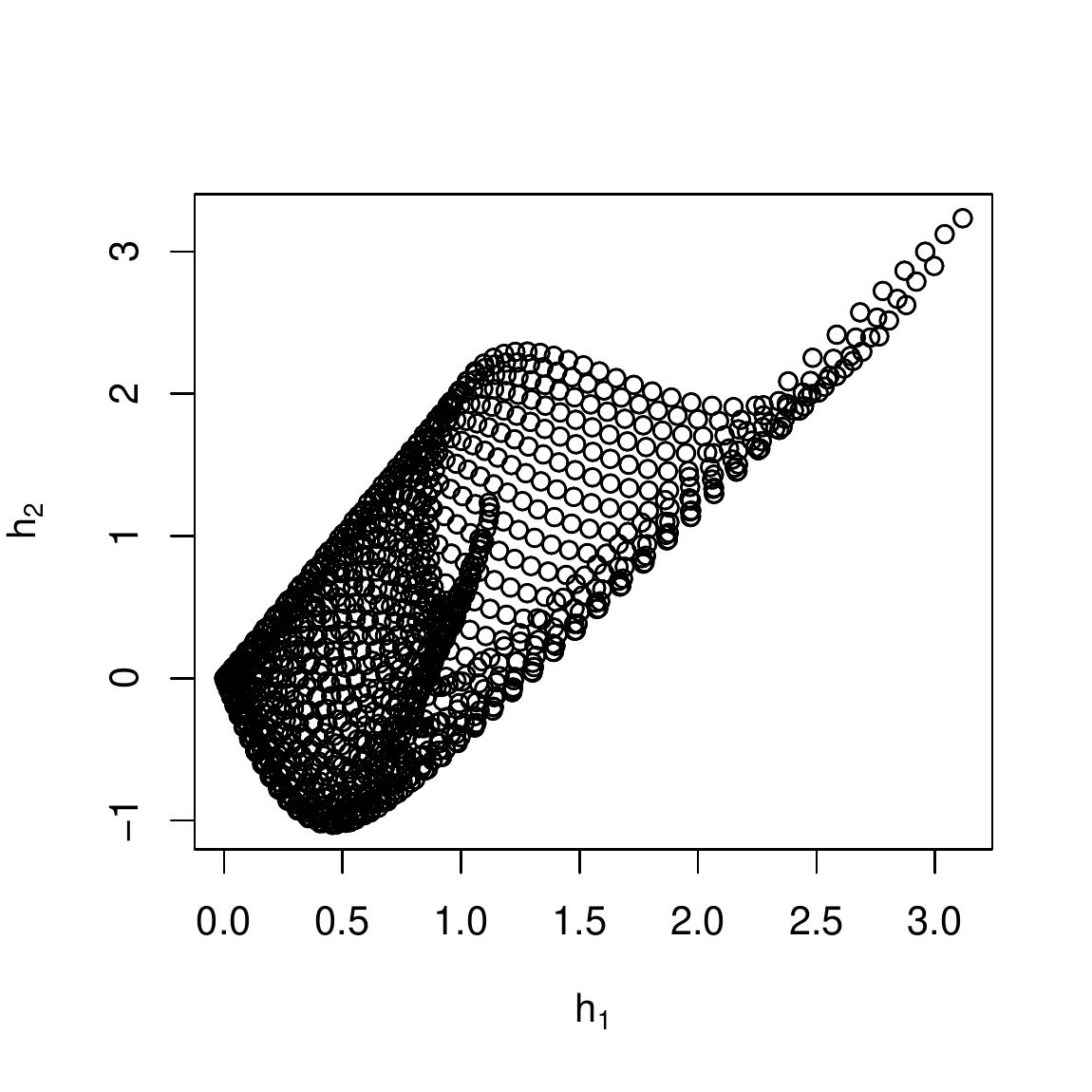}
\end{center}
\caption{ Outputs of the inner computer simulators.
 \label{Figure ex1-1}}
\end{figure}  
      
      Figure \ref{Figure ex1-1} illustrates the range of the inner computer outputs by conducting the inner computer experiments at $2000$ MmLHD points.  
   It can be seen that the inner computer outputs form a ``knife-like" shape. It is hard to construct an LHD in such an irregular design region. 
      Thus, the space-filling designs proposed by \cite{kyzyurova2018coupling}, which build space-filling designs for the inner and outer computer experiments separately, are not sufficient. 
      In addition, we can further see from Figure  \ref{Figure ex1-1} that more of the inner computer outputs are located at the positions where $h_1$ and $h_2$ are small.
      
\begin{figure}[htbp]
\begin{center}
\includegraphics[scale=0.8]{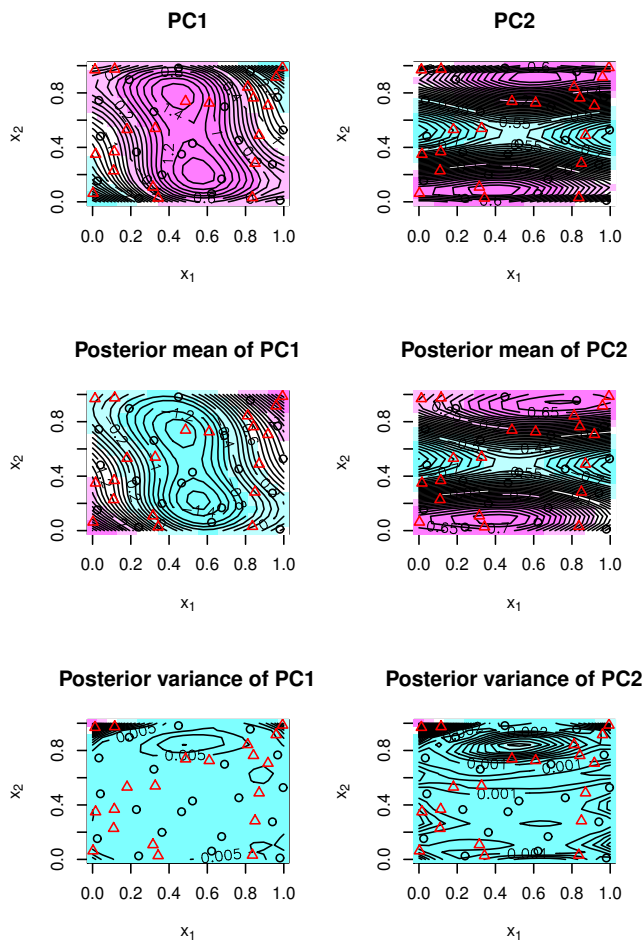}
\end{center}
\caption{The first row: the contour plots of ${\rm{PC}}_1$ and ${\rm{PC}}_2$. The second and the third row: the contour plots of the posterior mean  and posterior variance of ${\rm{PC}}_1$ and ${\rm{PC}}_2$, respectively. The red triangle signs are the sequentially added points  by the proposed method. 
 \label{Figure ex1-3}}
\end{figure}

Set $n_0=20$ and $N=2n_0$. By conducting PCA for the initial inner computer  outputs,  we have that the first PCs account for  $84.80\%$ of the total variation in the data, and  two PCs  account for  $100\%$ of the total variation. That is, $L^{pc}=L=2$. By choosing  Mat\'ern correlation functions (\ref{matern}) with smoothness parameter ${\nu}=5/2$ as  correlation functions, two GP models are built to mimic the two PCs. Following Algorithm \ref{alg:1},  Figure \ref{Figure ex1-3} shows  the relationship between the uncertainty of the GP models and the locations of the sequential points by the proposed method under the initial design ${\rm{ID}}_1$. The first row is the contour plots of ${\rm{PC}}_1$ and ${\rm{PC}}_2$. The second and the third row involve the contour plots of the posterior mean  and posterior variance of ${\rm{PC}}_1$ and ${\rm{PC}}_2$, respectively. The red plus signs are sequentially added points generated by the proposed method.
It can be seen that the sequentially added points are mainly concentrated in the places with the large absolute value of the derivatives and large posterior variance of the PCs.

{
The sequential designs obtained by three benchmark methods and the corresponding inner computer outputs are shown in Figure  \ref{Figure ex1-2}. 
  \begin{figure}[htbp]
\begin{center}
\includegraphics[scale=0.6]{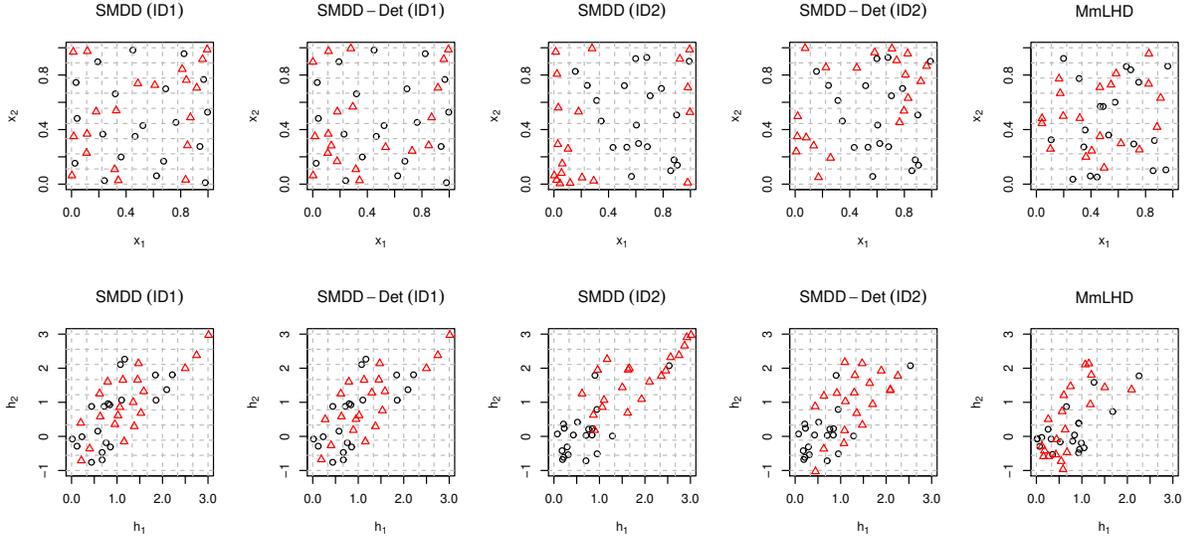}
\end{center}
\caption{ The sequential design points with $n_0=20$ and $N=40$. The black circles are the initial points and the red triangles are the  sequentially added points.  The ${\rm{ID}}_1$ in the first and the third columns refers to an initial MmLHD. The ${\rm{ID}}_2$ in the second and the fourth columns  indicate a poorly space-filled initial design, which doesn't explore the region $\{(x_1,x_2)\in[0,1]^2:x_1+x_2<0.5\}$. 
 \label{Figure ex1-2}}
\end{figure}  
By comparing the first column with others, we can see that the proposed method yields the most uniform design and the most uniform inner computer outputs. Specifically, because of ignoring  the posterior variance of PC1 and PC2, the sequential points obtained by the second method (SMDD-Det) are mainly concentrated in the places with the large absolute
value of the derivatives. Some locations  with large posterior variance, such as the point located around $(0.6,0.8)$ as shown in the first row and second column, are missed.  Moreover, the design and the corresponding computer outputs in the first column (in the second column) are more space-filled than that in the third column (in the fourth column). It indicates that the sequential design under the initial design ${\rm{ID}}_1$ outperforms the design under the initial design ${\rm{ID}}_2$. For the first method (SMDD), the poor initial design leads to large posterior variance at the unexplored region $\{(x_1,x_2)\in[0,1]^2:x_1+x_2<0.5\}$. Thus more sequential points are added at this region, resulting in poor spaced-filled inner computer outputs. For the second method (SMDD-Det), the low accuracy of the GP models leads to the region  $\{(h_1,h_2):h_1>2.5,h_2>2.5\}$ unexplored.  
The results in the fifth column show that the design and the corresponding inner computer outputs obtained by the last method have the poorest uniformity because $N-n$ MmLHD points are added randomly to the initial $n$ MmLHDs.
 }

\begin{figure}[htbp]
\begin{center}
\includegraphics[scale=0.6]{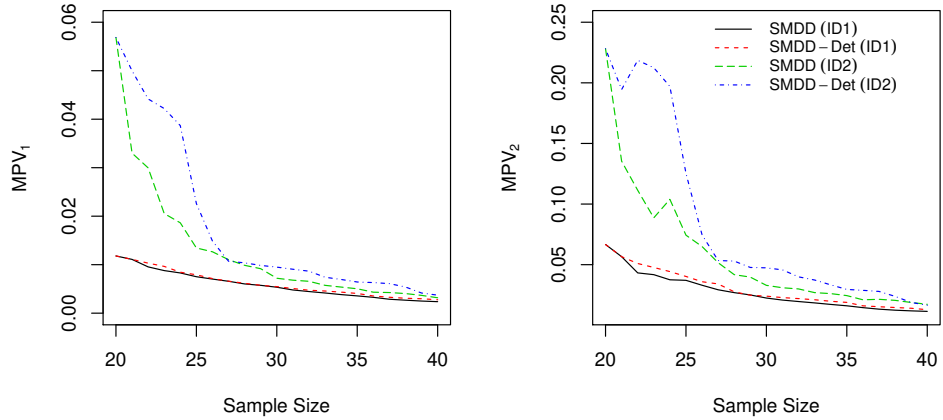}
\end{center}
\caption{MPV of the GP models for the two PCs v.s. the sample size as the number of sequential points increases.
 \label{Figure ex1-mse}}
\end{figure}

Denote the set of $\tilde N$ test points as  $\tilde {\bm X}=[\tilde{\bm x}_1,\ldots,\tilde{\bm x}_{\tilde{N}}]^T$, which is  a MmLHD. Let $\tilde N=500$, after adding $j,j=1,\ldots,N-n_0$ sequential points, 
the following  mean posterior variance (MPV) is used to evaluate the uncertainty of the GP models for the current PCs:
\begin{equation}
\label{mspe-outer}
 {\rm{MPV}}_l= \frac{1}{\tilde{N}}\sum_{i=1}^{\tilde{N}}s_l^2(\tilde{\bm x}_i), l=1,\ldots,L^{pc},
\end{equation}
where $s^2_l(\tilde{\bm x})$ (\ref{var}) is the posterior variance of   the GP predictor for the current $l$-th PCs at testing point $\tilde{\bm x}$.  Unlike the mean squared prediction error (MSPE), the MPV is only relevant to the location of $\tilde{\bm x}$. It is not necessary to run the computer experiment at  $\tilde{\bm x}$, so this criterion is also applicable to expensive computer experiments.
 Figure  \ref{Figure ex1-mse}  shows the variation of  mean posterior variance for the two PCs
 as the number of sequential points increases.



{It can be seen that the GP predictors for the two PCs under the  initial design ${\rm{ID}}_1$ are much more accurate than the predictors under the  initial design ${\rm{ID}}_2$. Since the proposed method considers the uncertainties of the GP predictors, the accuracy of the GP models under the proposed design is increased more quickly than that under the design given by the second method. }


{ 
  To further examine whether the proposed design can help to  improve the  prediction accuracy  of the outer surrogate model, we use the mean posterior variance (MPV) of the outer predictor to compare the accuracy of the outer surrogate model under different designs. 
  Two kinds of outer surrogate models, including a stationary GP model and a non-GP or a non-stationary GP model (here we use the Neural Network model),  are used to illustrate that the proposed design is robust to the model assumption for the outer computer simulators.
\begin{itemize}
\item GP model. 
Suppose the outer computer model is the modified Branin-Hoo function 
$$ g(h_1,h_2) = \left(h_2 - \frac{5.1}{4\pi^2}h_1^2 + \frac{5}{\pi}h_1 - 6\right)^2 + 10\left(1-\frac{1}{8\pi}\right)\cos(5 h_1) + 10,$$
$$(h_1,h_2)\in [-5,10]\times [0,15] .$$
 By choosing  the Mat\'ern correlation functions (\ref{matern}) with smoothness parameter ${\nu}=5/2$ as  correlation functions,  a GP model is used to mimic the Branin-Hoo function. 

\item Neural Network model. Suppose the outer computer model is a two-dimensional zigzag function 
    $$g(h_1,h_2)=h_1+2h_2-\floor{0.5+h_1}-\floor{0.4+h_2},h_1,h_2\in [0,1]^2.$$
    The zigzag is difficult to approximate by a stationary GP model because of its non-differentiable points \citep{lee2020neural}. With the help of the R package \emph{neuralnet} \citep{fritsch2016package},  a Neural Network model of depth $3$ is built to approximate the zigzag function.
\end{itemize}}

\begin{figure}[htbp]
\begin{center}
\includegraphics[scale=0.6]{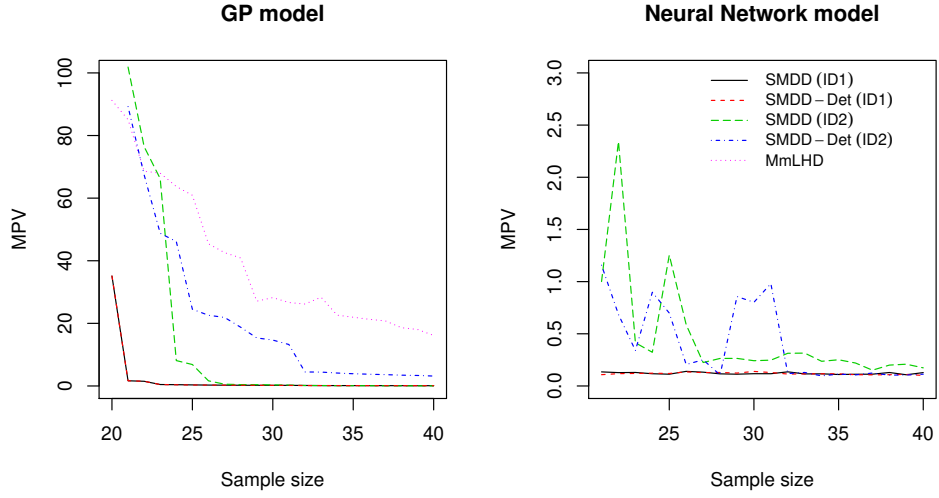}
\end{center}
\caption{MPV of the GP predictors (left) and MPV of the Neural Network predictors (right) v.s. the sample size as the number of sequential points increases.  (MPV of the Neural Network predictor under the last design is not shown because it is too large. )  
 \label{Figure ex1-mpv}}
\end{figure}

{
The test points are sampled from the 2000 MmLHD points as shown in  Figure \ref{Figure ex1-1}, so that the inner computer outputs at the testing points are space-filled in the ``knife-like"  region. 
Comparison of the accuracy of the GP predictors and the Neural Network predictors are shown in Figure  \ref{Figure ex1-mpv}.}
{
It can be seen that under the designs given by the proposed method, the GP and the Neural Network predictors for the outer computer simulators  perform best. It indicates that the proposed method is robust to the model assumption for the outer computer
simulators. The third method yields the worst-performing design. Moreover, the design given by the third method will lead  to a singular covariance matrix of the GP model as more sequential points are added.  It shows that the third method is not suitable for the design of nested computer experiments if the outer simulator is mimicked by a GP model.

}


{By repeating Algorithm \ref{alg:1} for $100$ times, Figure  \ref{Figure ex1-4} shows the mean of the AID of $\bm X_N$  and $\bm Y_I$ with different sample size $N$ of the  final sequential design.
  \begin{figure}[htbp]
\begin{center}
\includegraphics[scale=0.6]{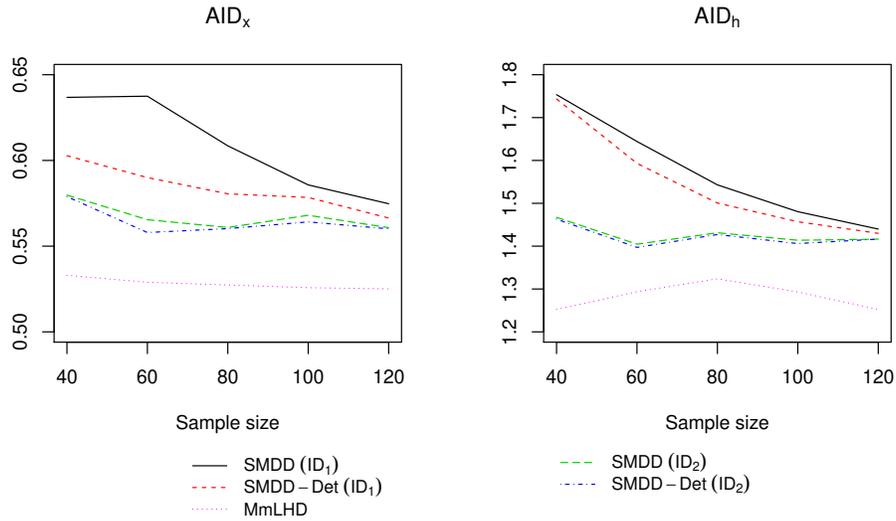}
\end{center}
\caption{The AID of $\bm X_N$ (left) and $\bm Y_I$ (right) v.s. the final sample size $N$, with fixed $n_0=20$ and $N_c=5(N-n_0)K=10(N-n_0)$.
 \label{Figure ex1-4}}
\end{figure}  
It can be seen that the proposed design outperforms all other methods when the sample size becomes large. 
From Figure \ref{Figure ex1-2}, we have that,  the sequential points are located at the boundary of the experimental space, such as the lower-left corner and the upper-right corner. It makes the ${\rm{AID}}_x$ for the sequential designs generated by the first two methods increase first and then decrease.
 With the further increase of the sample size, the ${\rm{AID}}_h$ will gradually decrease. 
By comparing the performance of the first two methods under the same initial design, we have that, under the initial design ${\rm{ID}}_1$, the proposed method achieves the largest average inter-site distance as the sample size becomes large. 
 The third method yields the worst-performing design.}

  \begin{example} [A high-dimensional example]   
      Suppose the outputs of the inner computer simulator at the point $\bm x$ are $10$ dimensional, and the $l$-th output is
   \begin{equation}
   \begin{aligned}
   h_l (\bm x)=&  4a_{l1} (x_1 - 2 + 8x_2 - 8x_2^2)^2+a_{l2}(3 - 4x_2)^2+
16 a_{l3}\sqrt{x_3+1}(2x_3-1)^2+\\
&a_{l4}\sum_{i=1}^8 i
\ln(1+\sum_{j=3}^i x_j), \bm x\in[0,1]^8,
   \end{aligned}
    \end{equation}
 where the parameters $a_{lj}, l=1,\ldots,10;j=1,\ldots,4$ is pre-fixed on $A=\left\{a_{lj}\right\}$ with
 \begin{equation}
 \begin{aligned}
 A^T
 =\begin{bmatrix}
0.614& 0.453& 0.264& 0.354 &0.850&0.514&0.040&0.958&0.142&0.717\\
 0.965& 0.400& 0.189& 0.574&0.323& 0.791& 0.093& 0.813&0.617& 0.221\\
 0.761& 0.410& 0.691& 0.872& 0.248& 0.574& 0.386& 0.086& 0.135& 0.938\\
 0.296& 0.189& 0.561& 0.775& 0.945& 0.002& 0.356& 0.615& 0.819& 0.435\\
 \end{bmatrix}.
 \end{aligned}
     \end{equation}
    \end{example}

\begin{figure}[htbp]
\begin{center}
\includegraphics[scale=0.55]{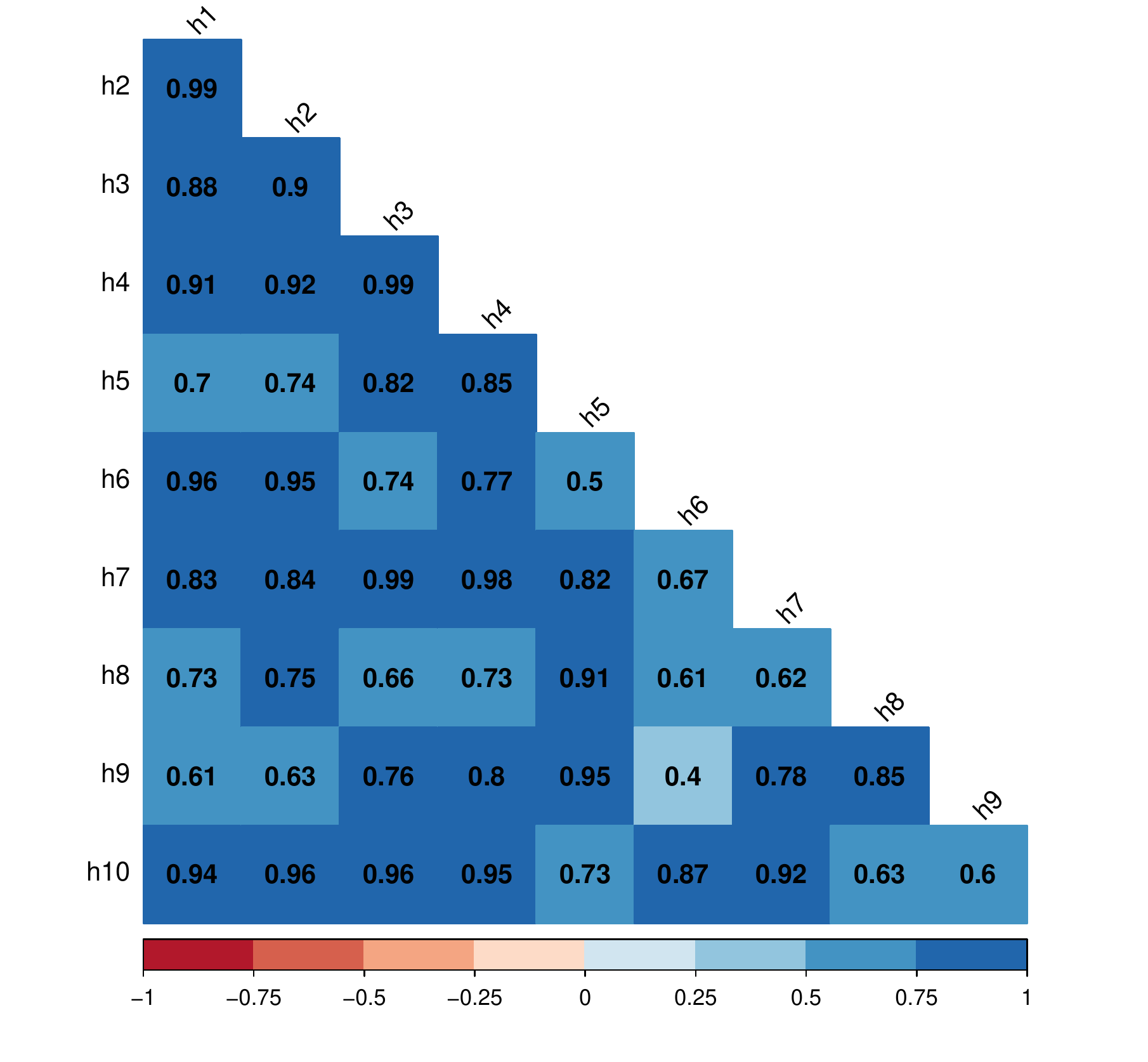}
\end{center}
\caption{The pairwise correlation of $\{\bm h_l\}_{l=1,\ldots, 10}$.
 \label{Figure ex2-1}}
\end{figure}

    Let $n_0=\max(10K,3L)=80$. The inner computer experiments are conducted at $n_0$ MmLHDs. The pairwise correlation between the $10$ vectors of inner computer outputs are shown in Figure \ref{Figure ex2-1}.
From the pairwise correlation plot (Figure \ref{Figure ex2-1}), we have that most of the inner computer outputs are highly correlated with the others. By performing SVD decomposition on the initial scaled $\bm Y_I$,  we have that the first two PCs
account for more than $90\%$ of the total variation in the data.

By choosing  Mat\'ern correlation functions (\ref{matern}) with smoothness parameter ${\nu}=5/2$ as  correlation functions, two GP models are built to mimic the PCs.
\begin{figure}[htbp]
\begin{center}
\includegraphics[scale=0.6]{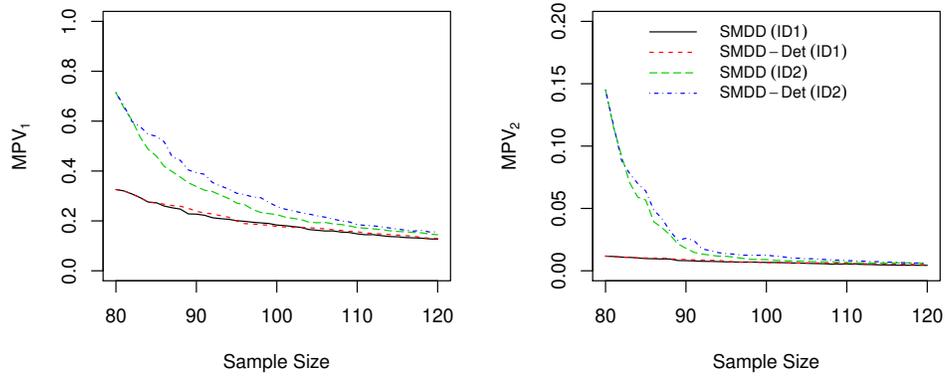}
\end{center}
\caption{ MPV of the GP models for the two PCs v.s. the sample size, as the number of sequential points increases.
 \label{Figure ex2-mse}}
\end{figure} 
Figure  \ref{Figure ex2-mse}  shows the variation of the MPV of the first two PCs  as the number of sequential points increases.
It can be seen that the GP models for the two PCs under the  initial design ${\rm{ID}}_1$ are much more accurate than the models under the  initial design ${\rm{ID}}_2$. As $40$ sequential points are added, the accuracy of the GP models under the design given by different methods tends to be the same.

{ 
Since the dimension of the inner computer output is $10$, building a non-stationary GP model or a non-Gaussian process model for the outer simulator is very time-consuming.
  We only  show the comparison results in the cases where the outer simulator can be approximated by a stationary GP model in this example.
Suppose the outer computer model is the wing weight function \citep{sobester2008engineering}. 
 By choosing  the Mat\'ern correlation functions (\ref{matern}) with smoothness parameter ${\nu}=5/2$ as  correlation functions,  a GP model is used to mimic this function. 
 A comparison of the accuracy of the GP predictors is shown in Figure  \ref{Figure ex2-mpv}.
\begin{figure}[htbp]
\begin{center}
\includegraphics[scale=0.65]{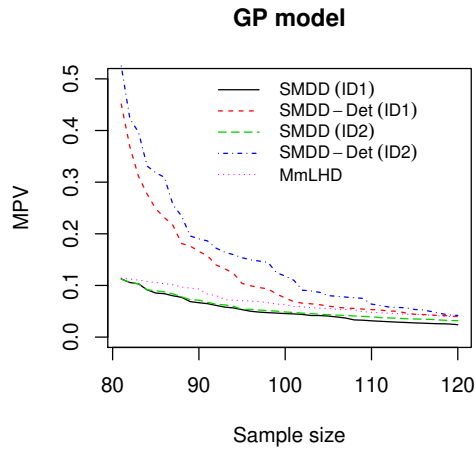}
\end{center}
\caption{MPV of the GP predictors v.s. the sample size as the number of sequential points increases.
 \label{Figure ex2-mpv}}
\end{figure}  
It can be seen that, under the designs given by the proposed method, the outer GP predictor performs best.
}

\begin{figure}[htbp]
\begin{center}
\includegraphics[scale=0.6]{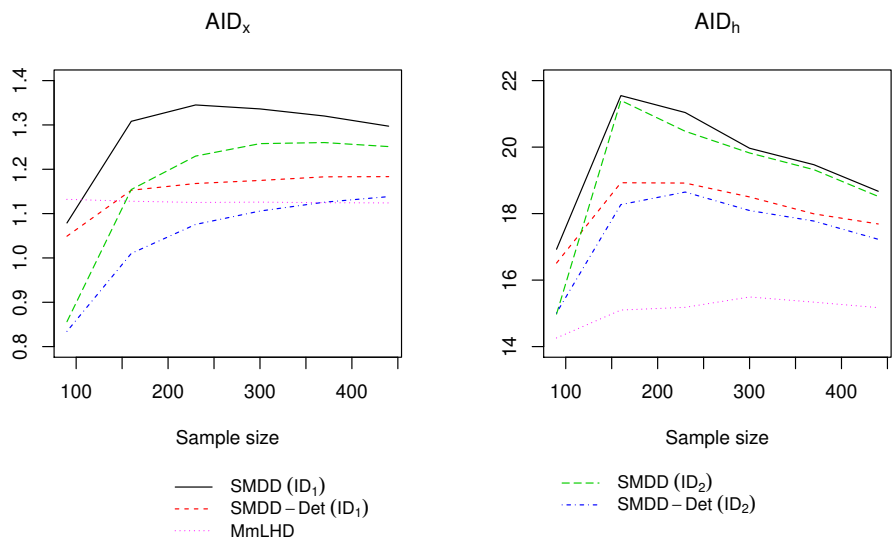}
\end{center}
\caption{ AID of $\bm X_N$  and $\bm Y_I$ v.s. the final sample size $N$ .
 \label{Figure ex2-2}}
\end{figure}

By repeating Algorithm \ref{alg:1} for $100$ times, Figure  \ref{Figure ex2-2} shows the mean of the AID of $\bm X_N$  and $\bm Y_I$ with different sample size $N$ of the  final sequential design.
{Due to the large dimension of $\bm x$, when the sample size increases by  $250$, ${\rm{AID}}_x$s of the first two methods still don't  have a very obvious decrease trend. For the third method, ${\rm{AID}}_x$s continue to decrease with the increase of sample size, although the decrease is not significant. The ${\rm{AID}}_h$ of the design given by the proposed method is larger than those of other methods, which illustrates the advantages of the proposed method.
 By comparing the first two methods under different initial designs, it can be seen that the poor initial design yields smaller AIDs. As the sample size increases, the difference from the initial design diminishes. By comparing the performance of the first two methods under the same initial design, we have that, under the initial design ${\rm{ID}}_1$, the proposed method achieves the largest average inter-site distance as the sample size becomes large. 
 The third method yields the smallest average inter-site distance for both $\bm X_N$ and $\bm Y_I$.  Combined with Figure \ref{Figure ex2-mpv}, we have  that designs with large AID lead to more accurate predictions. }

\section{Composite Structure Assembly Experiments}
\label{sec:case}
In this section, we apply the proposed design to the computer simulation for the composite structures assembly process.
As mentioned in Section \ref{sec:intro}, the composite structure assembly simulation involves two-layer computer models. The inner computer model is the Automatic Optimal Shape Control (AOSC) system that adjusts one fuselage (named part 1) to the target shape (assumed to be the shape of the other fuselage, named part 2) by ten actuators. After the shape adjustment, the two fuselages are assembled by the riveting process. Then, the actuators applied on the two fuselages will be released. It will cause the spring-back of the fuselages and the occurrence of residual stress. Residual stress may hurt the parts as well as generate other severe side
effects (e.g., fatigue, stress corrosion cracking, and structural instability)\citep{wen2019virtual}. Therefore,  the outer computer model is the platform that evaluates the residual stress during and after the assembly process.
Table     \ref{tab:com1} summarizes the inputs and outputs information in this computer experiment.

\begin{table*}[htbp]
    \centering
    \caption{Inputs and outputs for the  two-layer computer experiments}
    \label{tab:com1}
    \setlength{\tabcolsep}{5pt}
    \renewcommand{\arraystretch}{0.65}
 {  \begin{tabular}{lcccc}
        \hline
        Inner computer model\\
          &Name of variable& Dimension\\
          \hline
      Inputs&  Part 1’s actuators’ forces (${ \bm x}$)& $K=10$  \\
        Outputs&  Part 1’s critical dimensions  ($\bm h({ \bm x})$) & $L=53$  \\
               \hline
              Outer computer model\\
                     &Name of variable& Dimension &\\
          \hline
            Inputs&         Part 1’s critical dimensions  ($\bm h({ \bm x})$) & $L=53$& \\
            Outputs&  Maximum of the Stress & $M=1$  \\
                 \hline
    \end{tabular}}
   \end{table*}
 
\begin{figure}[htbp]
\begin{center}
\includegraphics[scale=0.6]{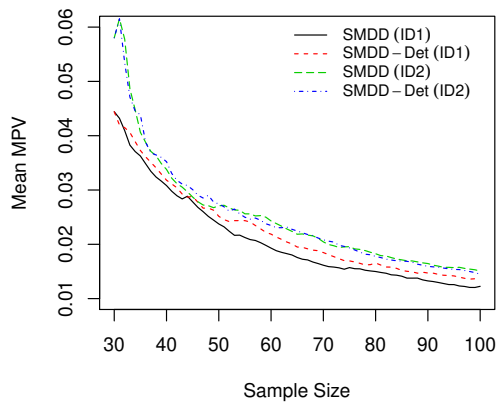}
\end{center}
\caption{Mean of the $L$ MPVs for the  independent GP models as the number of sequential points increases.
 \label{Figure case-mse-1}}
\end{figure}

Since the composite structure assembly experiments are  very time-consuming, it is assumed that the maximum sample size for the computer experiments is $100$ \citep{yue2018surrogate}.  Obviously, $3L$  is larger than $100$. It means that the sample size for the initial design cannot be set at $\max(10K,3L)$.
From \cite{yue2018surrogate}, building GP models for each $h_l,l=1,\ldots, 53$ independently  with $n_0^*=30$ training data can achieve satisfactory prediction performance. As a result, we ignore the correlation between the outputs. The PCA for the inner computer outputs $\bm Y_I$ can be omitted and only the accuracy of the GP models needs to be considered when choosing $n_0$.
Set $n_0=n_0^*=30$, and $N=100$. 
Figure \ref{Figure case-mse-1}    shows the mean of  $53$ MPVs as the number of sequential points increases.
It can be seen that the uncertainty of the $53$ GP models under the proposed design is smaller. As the sample size increases, the differences introduced by the different initial designs decrease. 

\begin{figure}[htbp]
\begin{center}
\includegraphics[scale=0.6]{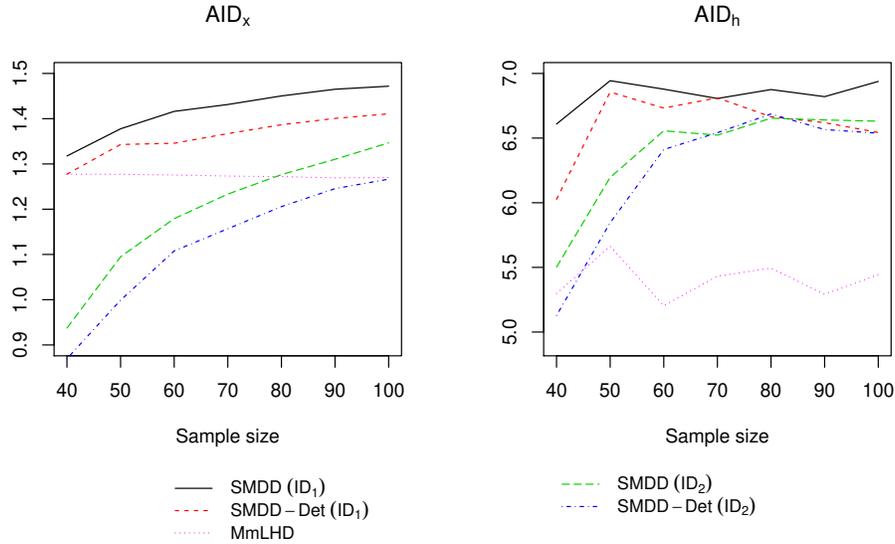}
\end{center}
\caption{ AID of $\bm X_N$  and $\bm Y_I$ v.s. the final sample size $N$.
 \label{Figure case-2}}
\end{figure}

Due to the impact of arithmetic power, we only show the average inter-site distance obtained by running Algorithm \ref{alg:1} once.
Figure  \ref{Figure case-2} shows the mean of the AIDs of $\bm X_N$  and $\bm Y_I$ with different sample size $N$ of the  final sequential design.
Due to the high dimensionality  of $\bm x$ and $\bm h$,  AIDs for the first two methods don't have a significant decreasing trend with $N=100$. 
The proposed design outperforms all other methods when the sample size becomes large. 
 By comparing the first two methods under different initial designs, it can be seen that when the sample size is small, ${\rm{AID}}_h$s under the second initial design are much smaller, due to the larger uncertainty of the GP models. By comparing the performance of the first two methods under the same initial design, we have that, under the initial design ${\rm{ID}}_1$, the proposed method achieves the largest average inter-site distance as the sample size becomes large. 
 The third method yields the worst-performing design.

\section{Conclusion and Discussion}
\label{sec.dis}

In this work, we proposed a  sequential design procedure for the  two-layer computer experiments.  The proposed design considers the space-filling property of both the inputs and the outputs for the inner computer experiments. A new sequential algorithm for efficiently generating such designs is also given. 
The numerical simulation results show the proposed design out-performance of the benchmark designs.  At last, the proposed method is applied to generate designs for the composite structures assembly experiments.

 The proposed method is robust to the assumption for the outer computer simulators but relies on the accuracy of the surrogate model for the inner computer simulator.
An inner computer model-independent design approach will be considered later.

\section*{Supplementary Material}
The data in case study and the corresponding R codes are available which are provided in the Supplementary Material. All files consist the .zip file.

\begin{appendices}

\section{Gaussian Process models}
\label{emu-gp}

In this section, we briefly introduce the GP modeling for the $l$-th PCs, $l=1,\ldots, L^{pc}$.
Suppose $ h_l^{pc}(\cdot)$ is one realization of the GP
\begin{equation}
\begin{aligned}
 h_l^{pc}(\cdot)&= \bm\beta^T \bm b(\cdot)+Z(\cdot),\\
  Z(\cdot)| \theta&\sim GP( {0},  \sigma^2R(\cdot,\cdot)).
\end{aligned}
\end{equation}
$\bm b(\cdot)$ consists of $q$ basis functions; $\bm\beta$ is the corresponding  regression coefficients; $GP( 0,\sigma^2 R)$ denotes a stationary GP with mean zero,   covariance function $\sigma^2 R(\cdot,\cdot)$. Here,   $\sigma^2>0$ is the variance and $R(\cdot,\cdot)$ is the  correlation function.
Common choices of $R(\cdot,\cdot)$ include the Gaussian correlation functions with
 \begin{eqnarray}
R(\bm{x},\bm{x}')=\exp(-\theta {\parallel \bm{x}-\bm{x}'\parallel}^2),
\label{exp}
\end{eqnarray}
and the Mat\'ern correlation functions with
 \begin{eqnarray}
R(\bm{x},\bm{x}')=\frac{1}{\Gamma(\nu)}\left(\frac{2\sqrt{\nu}\parallel \bm{x}_i-\bm{x}_j\parallel}{\theta}\right)^{\nu}K_{\nu}\left(\frac{2\sqrt{\nu}\parallel \bm{x}-\-\bm{x}'\parallel}{\theta}\right),
\label{matern}
\end{eqnarray}
where $\theta>0$ is the correlation parameter and  $K_{\nu}$ denotes the modified Bessel function of the second kind with order ${\nu}$.

Recall the inner computer experiments are conducted at the points $\bm X_n=(\bm x_1,\ldots,\bm x_n)^T$ and the corresponding outputs of $h_l^{pc}$ are   $\bm H^{pc}_l=[h^{pc}_l(\bm x_1),\ldots,h^{pc}_l(\bm x_n) ]^T$.  
Denote ${{ {\bm R}}}= R(\bm X_n, \bm X_n)$;  ${\bm r}(\bm x)= R(\bm x, \bm X_{n})$ and $\mathbf{B}^T=\left[ \bm b(\bm x_1),\ldots, \bm b(\bm x_n)\right]$. Given data $(\bm X_n,\bm H^{pc}_l)$, the posterior distribution of $h_l^{pc}(\bm x)$ is
\begin{equation}
\label{posnest}
h_l^{pc}( \bm x)|\bm X_n,\bm H^{pc}_l \sim N\left( {\hat h_l^{pc}}(\bm  x),s_l^2(\bm x)\right).
\end{equation}
Here, the posterior mean is
\begin{equation}
\label{blue}
 {\hat h_l^{pc}}(\bm  x)=\hat{\bm\beta}^T \bm b(\bm x )+(\mathbf Y_I-\mathbf{B}\hat{\bm\beta})^T{\bm{R}}^{-1}{\bm r}(\bm x ),
\end{equation}
and the posterior variance is 
\begin{equation}
\begin{aligned}
\label{var}
s_l^{2}(\bm x)&=\hat\sigma^2 \tilde R(\bm x,\bm x),\\
\end{aligned}
\end{equation}
with
$$\tilde R(\bm x,\bm x')=R(\bm x ,\bm x' )-{\bm r}^T(\bm x ){\bm{R}}^{-1}{\bm r}(\bm  x')+\bm U^T(\bm x )(\mathbf{B}^T{\bm{R}}^{-1}\mathbf{B})^{-1}\bm U(\bm x' ),$$
and 
$$\hat\sigma^2 =(n-1)^{-1}(\bm H^{pc}_l-\bm B\hat{\bm\beta})^T\bm {R}^{-1}(\bm H^{pc}_l-\bm B\hat{\bm\beta}).$$ 
Here, $\hat{\bm\beta}=(\mathbf{B}^T{\bm{R}}^{-1}\mathbf{B})^{-1}\mathbf{B}^T{\bm{R}}^{-1}\bm H^{pc}_l$ and
  $\bm U(\bm x )=\bm b(\bm x )-\mathbf{B}^T{\bm{R}}^{-1}{\bm r}(\bm x )$.
 In addition,   the hyper-parameter $\theta$ in the correlation function is always unknown in practice, maximum likelihood estimators (MLEs)  can be plugged into (\ref{posnest}) to obtain the posterior distribution of $h_k$.

\end{appendices}
\bibliographystyle{chicago}

\bibliography{boref}

\end{document}